\documentstyle[preprint,aps,prl,epsf]{revtex}
\input{epsf}
\input epsf.sty

\begin{document}
\draft
\title{COHERENCE IN THE QUASI-PARTICLE 'SCATTERING' BY THE VORTEX 
LATTICE IN PURE
TYPE-II SUPERCONDUCTORS}
\author{V.N.Zhuravlev and T.Maniv}
\address{Department of Chemistry, Technion-Israel Institute of 
Technology\\
Haifa 32000, ISRAEL 
.}
\author{I.D.Vagner and P.Wyder}
%EndAName
\address{Grenoble High Magnetic Field Laboratory \\
Max-Planck-Institute f\"ur Festkorperforschung and\\
Center National de la Recherche Scientific,\\
\ \ 25 Avenue des Martyres, F-38042, Cedex 9, FRANCE \\
%EndAName
.}
\date{\today}
\maketitle

\begin{abstract}
The effect of quasi-particle (QP) 'scattering' by the vortex lattice on the
de-Haas van-Alphen oscillations in a pure type-II superconductor is
investigated within mean field,asymptotic perturbation theory. Using a 2D
electron gas model it is shown that, due to a strict phase coherence in the
many-particle correlation functions, the 'scattering' effect in the
asymptotic limit ($\sqrt{E_F/\hbar\omega_c}\gg 1$) is much weaker than what
is predicted by the random vortex lattice model proposed by Maki and
Stephen, which destroys this coherence . The coherent many particle
configuration is a collinear array of many particle coordinates , localized
within a spatial region with size of the order of the magnetic length. The
amplitude of the magnetization oscillations is sharply damped just below $%
H_{c2}$ because of strong $180^{\circ}$ out of phase magnetic oscillations
in the superconducting condensation energy ,which tend to cancel the normal
electron oscillations. Within the ideal 2D model used it is found, however,
that because of the relative smallness of the quartic and higher order terms
in the expansion , the oscillations amplitude at lower fields does not
really damp to zero, but only reverses sign and remains virtually undamped
well below $H_{c2}$. This conclusion may be changed if disorder in the
vortex lattice, or vortex lines motion will be taken into account. The
reduced QP 'scattering' effect may be responsible for the apparent crossover
from a strong damping of the dHvA oscillations just below $H_{c2}$ to a
weaker damping at lower fields observed experimentally in several 3D
superconductors.

\end{abstract}

\section{Introduction}

Magnetic Quantum Oscillations have been recently observed in several type-II
superconductors below $H_{c2}$ \cite{onu92}-\cite{wel95}. A systematic study
of this remarkable effect has been impaired, however, by the lack of a
complete quantitative theory of the de Haas-van Alphen (dHvA) effect in the
vortex state, analogous to the Lifshitz-Kosevich (LK) theory in normal
metals \cite{lif56}. Such a theory would require a detailed analysis of the
effect of the superconducting order parameter on the magnetization
oscillations in the vortex state, which turns out to be an extremely subtle
theoretical problem.

A common feature reported by all experimental groups so far,which is far
from being well understood, has been the observation of an additional
damping in the dHvA amplitude below $H_{c2}$. Several theoretical papers
have attributed this attenuation to the broadening of the Landau levels by
the inhomogeneous pair potential, building up below $H_{c2}$. There is,
however, a remarkable disagreement among the various theoretical approaches
to this problem concerning both the size of the attenuation factor and its
detailed dependence on the strength of the field below $H_{c2}$.

The semiclassical approach, adopted originally by Maki \cite{mak91},
elaborated later by Stephen \cite{ste92}, and reviewed very recently by
Wasserman and Springford \cite{wass96}, considered the correction to the
Quasi-Particles (QP) lifetime due to the 'scattering' by the vortex lattice.

For the first harmonic of the oscillatory part of the magnetization below $%
H_{c2}$ they predicted: 
\begin{equation}
M_{osc}=M_{n,osc}\exp \left[ -\lambda\left( \Delta _0,n_F\right) \right]
\label{p1.1}
\end{equation}
with 
\begin{equation}
\lambda\left( \Delta _0,n_F\right) =\pi ^{3/2} \frac{\tilde \Delta _0^2} {%
n_F^{1/2}}  \label{p1.2}
\end{equation}
where $M_{n,osc}$ is the corresponding normal electrons contribution to the
oscillatory magnetization, $\tilde \Delta _0\equiv \Delta_0 /\hbar \omega _c$%
, $\Delta _0$ - the magnitude of the superconducting (SC) order parameter, $%
\omega_c$-the cyclotron frequency, and $n_F=E_F/\hbar \omega _c$ is the
Landau level index corresponding to the extremal orbit on the Fermi surface.

An exponential damping with a different exponent, i.e. $\lambda\sim\tilde{
\Delta}_{0}/n_{F}^{1/4}$, has been proposed by Norman et al.\cite{nor95},
who carried out a full quantum mechanical calculation, based on a numerical
solution of the Bogoliubov-de Gennes (BdG) equations for the quasi particles
at low temperatures. The numerical computations carried out by these authors
were limited, however, to relatively small values of $\sqrt{n_F}$.

A similar approach, invoked by Dukan and Tesanovic \cite{duk95}, has led to
a qualitatively different behavior at very low temperature, that is a power
law attenuation of the dHvA oscillations below $H_{c2}$ 
\begin{equation}
M_{osc}/M_{n,osc}\sim \left( k_BT/\Delta _0\right) ^2+{\it O}\left[ \left(
k_BT/\Delta _0\right) ^4\right]  \label{p1.4}
\end{equation}

This result was obtained by considering only the dominant contribution to
the dHvA oscillations as originating in the 'gapless' region of the QP
spectrum around the Fermi surface. It looks like a low temperature high $%
\Delta $ expansion, emphasizing the opening of the SC gap well below $H_{c2}$

In the Maki-Stephen (MS) theory it is assumed that the vortex lattice acts
like a random potential for the quasi particles, and so by averaging over
the realizations of the vortex lattice, the QP self energy acquires a large
imaginary part, leading to a strong exponential damping of the dHvA
amplitude.

Strictly speaking, however, the 'scattering' by the inhomogeneous pair
potential is a highly coherent process, as in multiple Andreev reflection at
the interfaces of a 2D periodic array of normal and SC phases \cite{klein93}.

Thus, in the ideal, self consistent vortex lattice model the QP self energy
has zero imaginary part. However, the broadening of the Landau levels into
real energy bands by the inhomogeneous pair potential should lead to damping
of the dHvA oscillations even for an infinite QP 'lifetime'. The term
'scattering' refers to this inhomogeneous broadening effect in the present
paper.

The quadratic dependence of $\lambda(\Delta _0,n_F)$ on $\Delta _0$ in Eq.( 
\ref{p1.2}) reflects its origin in a perturbation expansion of the the QP
self energy in the mean square order parameter, which is strictly valid for
a pure type-II superconductor only at sufficiently high temperatures, when
the Lifshitz-Kosevich \cite{lif56} thermal smearing factor $X\equiv 2\pi^2
k_BT/\hbar \omega _c>1$.

It is therefore very interesting to compare the result expressed in Eq.(\ref
{p1.1}) to that obtained by Maniv et al.\cite{man92,man94}, who considered
Gorkov's expansion of the SC free energy in the small vortex state order
parameter near $H_{c2}$.

Using a semiclassical approximation, valid for $\sqrt{n_F}\gg 1$, these
authors have found a quadratic term consistent with Eq.(\ref{p1.2}), but a
quartic term $M_{osc}^{(4)}/M_{n,osc}\sim \tilde{\Delta}_{0}^{4}/n_{F}^{3/2}$
, which is smaller than the corresponding term, obtained in MS theory (i.e. $%
M_{osc}^{(4)}/M_{n,osc}\sim \lambda^2 \sim \tilde{\Delta}_{0}^{4}/n_{F}$),
by the factor $1/\sqrt{n_{F}}\ll 1$.

Very recently, Bruun et al. \cite{bruun97} developed an exact numerical
scheme for calculating the coefficients of the Gorkov's expansion within the
same model used by Maniv et al. and found good agreement with the results
obtained by Norman et al. for small values of $n_F$. They have also made an
estimate of the $n_F$ dependence of the quartic term by using an
approximation similar in spirit to the random lattice approximation, and
found a result which agrees with that of Maki and Stephen.

In the light of this controversy our purpose in the present paper is to
carefully examine the high temperature $X\geq 1$, small $\Delta _0$,
asymptotic ($\sqrt{n_F}\gg 1$) expansion, in order to elucidate the origin
of the disputed $n_F$ dependence. We find that, incoherent 'scattering'
channels, which generate the dominant contribution in the random lattice
approximation, are completely cancelled in the self-consistent , periodic
lattice calculation, due to the presence of a strict phase coherence in the
four-particle correlation function. The remaining coherent four-particle
configuration is a collinear array of four-electron coordinates, localized
within a spatial region with size of the order of the magnetic length.

As a result, the inhomogeneous broadening of the Landau levels by the
pair-potential does not contribute significantly to the damping of the dHvA
amplitude just below $H_{c2}$ , as is the case in the MS theory. The
dominant damping mechanism in the asymptotic limit ,$n_F^{1/2}\gg 1$ is
found to arise from the strong, $180^{\circ}$ out of phase oscillations of
the SC condensation energy with respect to the normal electrons
oscillations, as was first proposed in Ref.\cite{man94}.

The organization of the paper is as follows: In Sec.II we present the
general framework of our approach, which is based on the Gorkov's expansion
of the SC free energy near $H_{c2}$. In Sec.III we carefully examine the
quartic term in the asymptotic limit $\sqrt{n_{F}}\gg 1$ and justify the
main approximation used in our calculation. In Sec.IV we derive simple
analytical expressions for the self consistent order parameter and for the
oscillatory magnetization , and verify the validity of our claimed $n_F$
dependence in the asymptotic limit. In Sec.V we discuss the connection
of our theory to the other theoretical approaches and compare our 
predictions to experiment.

\section{Small Order Parameter Expansion}

We consider the quadratic $F_s^{\left( 2\right) }$ and the quartic $%
F_s^{\left( 4\right) }$ terms in the Gorkov-Ginsburg-Landau expansion of the
SC free energy in the SC order parameter $\Delta \left( \overrightarrow{r}%
\right) $ \cite{man92}. For the sake of simplicity let us assume a 2D
electron gas model and neglect the spin degrees of freedom. The former
assumption may be justified by noting that the main contribution to the dHvA
effect in an isotropic 3D normal electron system comes from the extremal
orbit, corresponding to the value $k_z =0$ of the electron momentum parallel
to the field direction, and to the Landau level $n\approx n_F\equiv \frac{E_F%
}{\hbar\omega_c}$.

Note, however, that the pairing effect responsible for the Cooper 
instability in a 3D electron gas is dominated by the region 
$k_z\approx k_F\equiv (2mE_F/\hbar^2)^{1/2}$, $n\approx 0$ near the Fermi
surface. Since the focus in the present paper is on the effect of QP
'scattering' by the vortex lattice, we shall ignore, for the sake of
simplicity, this aspect of 3D systems in our calculation, but will return to
this problem later while discussing our results in connection with
experiment.

Thus 
\begin{equation}
F_s^{\left( 2\right) }=\frac 1V\int d^2r\left|\Delta\left(\overrightarrow{ r}%
\right) \right| ^2-\int d^2r_1d^2r_2K_2\left( \overrightarrow{r}_1,%
\overrightarrow{r}_2\right) \Delta \left( \overrightarrow{r}_1\right) \Delta
^{\star }\left( \overrightarrow{r}_2\right)  \label{p2.1}
\end{equation}
and 
\begin{equation}
F_s^{\left( 4\right) }=\frac 12\int d^2r_1d^2r_2d^2r_3d^2r_4K_4\left(
\left\{ \overrightarrow{r}_i\right\} \right) \Delta \left( \overrightarrow{ r%
}_1\right) \Delta ^{\star }\left( \overrightarrow{r}_2\right) \Delta \left( 
\overrightarrow{r}_3\right) \Delta ^{\star }\left( \overrightarrow{r}%
_4\right)  \label{p2.2}
\end{equation}
where $V$ is BCS interaction constant. The kernels $K_2$ and $K_4$ are
expressed through the product of two and four electron Green's functions $%
G_0\left( \overrightarrow{r}_l,\overrightarrow{r}_{l+1};\left( -\right)
^l\omega _\nu \right) $, $l=1,2$ with $\overrightarrow{r}_3\equiv 
\overrightarrow{r}_1$, and $l=1,...,4$ with $\overrightarrow{r}_5\equiv 
\overrightarrow{r}_1$, respectively, in magnetic field (in our case$%
\overrightarrow{H}=\left( 0,0,H\right) $). Here $\omega _\nu $ is the
Matzubara frequency. The magnetic field breaks translational symmetry of BCS
Hamiltonian and Green's function. However, owing to the gauge symmetry the
Green's function can be represented in a factorized form: $G_0\left( 
\overrightarrow{r}_l,\overrightarrow{r}_{l+1};\omega _\nu \right) =g\left( 
\overrightarrow{r}_l,\overrightarrow{r}_{l+1}\right) \widetilde{G}_0\left(
\rho _l,\omega _\nu \right) $ \cite{byc62}, where the reduced Green's
function 
\begin{equation}
\widetilde{G}_0\left( \rho _l,\omega _\nu \right) =\frac 1{2\pi a_H^2}%
e^{-\rho _l^2/4}\sum\limits_n\frac{L_n\left( \rho _l^2/2\right) } {i\omega
_\nu -\omega _c\left( n+1/2\right) +\mu }  \label{p2.2a}
\end{equation}
depends only on relative coordinates $\overrightarrow{\rho }_l=\left( 
\overrightarrow{r}_{l+1}-\overrightarrow{r}_l\right) /a_H$, $\sum 
\overrightarrow{\rho }_l\equiv 0$, $a_H=\sqrt{\frac{c\hbar }{eH}}$. In this
expression $L_n$ is the Lagaurre polynomial of order $n$, and $\mu $ is the
chemical potential. The non invariant gauge factor in the symmetric gauge
has the form: $g\left( \overrightarrow{r}_l,\overrightarrow{r}_{l+1}\right)
=\exp \left( -i\varepsilon _{ik}\left( r_{l,i}+r_{l+1,i}\right) \rho
_{l,k}/4a_H\right) $, $\varepsilon _{lk}=-\varepsilon _{kl}$ is
antisymmetric tensor in 2D space.

In quasiclassical limit ( $\sqrt{n_F}\gg 1$ ) Green's function $\widetilde{G}%
_0\left( \rho _l,\omega _\nu \right) $ has two different types of behavior.
Near turning point $\rho _l\simeq 2r_F$ ( $r_F=\sqrt{2n_F}$ is the cyclotron
radius ) it is a smooth function, and in the intermediate region $\rho
_l<2r_F$ it is a sharply oscillating function.

From now on we shall express all spatial variables, except for $%
\overrightarrow r_{i}$, in units of the magnetic length.

Using asymptotic for Lagaurre polynomial $L_n$ at $n\rightarrow \infty $ 
\cite{bat53} and the Poisson summation formula for the sum over $n$ in (\ref
{p2.2a}) one can show that $\widetilde{G}_0\left( \rho _l,\omega _\nu
\right) \sim 1/n_F^{1/3}$ at $4r_F^2-\rho _l^2\leq \left( 16n_F/3\right)
^{1/3}$, and 
\begin{equation}
\widetilde{G}_0\left( \rho _l,\omega _\nu \right) =-\frac{i\varepsilon
_{\omega _\nu }J\left( \omega _\nu \right) }{\left( 2\pi \right)
^{1/2}a_H^2\omega _c}\frac{\exp \left[ i\varepsilon _{\omega _\nu }\left[
n_F\left( \phi +\sin \phi \right)\right] -\phi \frac{\left| \omega _\nu
\right| }{\omega _c} \right] }{\rho _l^{1/2}\left( r_F^2-\rho _l^2\right)
^{1/4}}  \label{p2.2b}
\end{equation}
where $J^{-1}\left( \omega _\nu \right) =1-\exp \left( 2\pi i\varepsilon
_{\omega _\nu }n_F-2\pi \left| \omega _\nu \right| /\omega _c\right) $, sin$%
\phi =\frac{\rho _l}{r_F}\left( 1-\left( \frac{\rho _l}{2r_F}\right)
^2\right) ^{1/2}$, $\varepsilon _{\omega _\nu }=sign\left( \omega _\nu
\right) $, at $\rho _l<2r_F$. Note that this expression is a generalization
of the Green's function obtained in \cite{man92} for $\rho _l/2r_F\ll 1$.
The similarity to the short distance limit is apparent since the analytic
continuation of the Lagaurre polynomial asymptotic in the complex $n$ plane
gives rise to short distance approximation at $\left| n\right| \rightarrow
\infty $.

Combining the gauge factors we obtain the dependence of $K_2$ and $K_4$ on
the two and four-particle center of mass coordinates $\overrightarrow{r}%
=\left( \sum_{l=1}^2\overrightarrow{r}_l\right) /2a_H$ and $\overrightarrow{R%
}=\left(\sum_{l=1}^4\overrightarrow{r}_l\right) /4a_H$ respectively: 
\begin{equation}
K_2\left( \overrightarrow{r}_1,\overrightarrow{r}_2\right) =\exp \left(
-i\varepsilon _{lk}r_l\rho _k\right) \frac 1\beta \sum_\nu \widetilde{K}%
_{2,\nu }\left( \rho \right)  \label{p2.3}
\end{equation}
\begin{equation}
K_4\left( \left\{ \overrightarrow{r}_i\right\} \right) =\exp \left(
-4i\varepsilon _{lk}R_lQ_k\right) \frac 1\beta \sum_\nu \widetilde{K}_{4,\nu
}\left( \left\{ \rho _l\right\} \right)  \label{p2.4}
\end{equation}
Where $\overrightarrow{\rho }\equiv \overrightarrow{\rho }_1$, $%
\overrightarrow{Q}=-\left( \overrightarrow{\rho }_1-\overrightarrow{\rho}_2+ 
\overrightarrow{\rho }_3-\overrightarrow{\rho }_4\right) /8$ and 
\begin{equation}
\widetilde{K}_{2,\nu }\left( \rho \right) =\widetilde{G}_0\left( \rho
,-\omega _\nu \right) \widetilde{G}_0\left( \rho ,\omega _\nu \right)
\label{p2.5}
\end{equation}
\begin{equation}
\widetilde{K}_{4,\nu }\left( \left\{ \rho _l\right\} \right) =\widetilde{G}%
_0\left( \rho _1,-\omega _\nu \right) \widetilde{G}_0\left( \rho _2,\omega
_\nu \right) \widetilde{G}_0\left( \rho _3,-\omega _\nu \right) \widetilde{G}%
_0\left( \rho _4,\omega _\nu \right)  \label{p2.6}
\end{equation}

For the order parameter we use the Abrikosov form. In symmetric gauge 
\begin{equation}
\Delta \left( \overrightarrow{r}\right) =\Delta _0\exp \left(
-y^2+ixy\right) \sum_{n=-\infty }^\infty \exp \left[ -\left( 1-i\gamma
\right) \left( \frac{\pi n}{a_x}\right) ^2+2i\frac{\pi n}{a_x}\left(
x+iy\right) \right]  \label{p2.6a}
\end{equation}
The numbers $\gamma $ and $a_x$ are arbitrary. It can be shown that this
quasiperiodical form results from gauge symmetry as a solution minimizing
free energy \cite{man92}. The order parameter is normalized as 
\begin{equation}
\Delta _0^2=\frac{\sqrt{2/\pi}}{a_x a_H^2 N}\int d^2r\left| \Delta \left( 
\overrightarrow{r}\right) \right| ^2  \label{p2.6b}
\end{equation}
where $N$ is the number of vortices.

Using expression (\ref{p2.6a}) for the order parameter in Eq.(\ref{p2.1})
for the quadratic term we get after integrating over the two-particle center
of mass coordinate $\overrightarrow{r}$

\begin{equation}
F_s^{\left( 2\right) }=\left( \frac 1V-{\cal A}\right) \pi a_H^2N\Delta _0^2
\label{p2.7}
\end{equation}
where ${\cal A}\propto \frac 1\beta \sum\limits_\nu \int d^2\rho e^{-\rho
^2/2}\widetilde{K}_{2,\nu }\left( \rho \right) $. The gaussian factor
restricts the effective integration region by the distances $\rho \sim 1$.
This important fact means the loss of coherence in electron pair propagation
over distances much larger than the magnetic length.

The localization of the electron correlation functions within a region of a
size of the order of the order of the magnetic length can be shown to exist,
in quasiclassical limit, also for many electron configurations. Here we
discuss only four particle correlations. Integration of (\ref{p2.2}) over $%
\overrightarrow{R}$ gives rise to expression 
\begin{equation}
F_s^{\left( 4\right) }=\int d^2Qe^{-4Q^2}\int d^2Sd^2T{\cal D}\left( 
\overrightarrow{S},\overrightarrow{T}\right) \frac 1\beta \sum_\nu 
\widetilde{K}_{4,\nu }\left( \left\{ \rho _l\right\} \right)  \label{p2.8}
\end{equation}
where $\overrightarrow{S}\equiv \frac 14\left( \overrightarrow{\rho }_3-%
\overrightarrow{\rho }_1\right) $, $\overrightarrow{T}\equiv \frac 14\left( 
\overrightarrow{\rho }_4-\overrightarrow{\rho }_2\right) $. The function $%
{\cal D}\left( \overrightarrow{S},\overrightarrow{T}\right) $ includes the
summation over vortices: 
\[
{\cal D}\left( \overrightarrow{S},\overrightarrow{T}\right) \propto
Ne^{4i\left( S_xT_y+S_yT_x\right) }\sum_{m_1,m_2}e^{-2i\gamma
m_1m_2+4im_1T_x+4im_2S_x} 
\]
\begin{equation}
\exp \left[ -\left( m_1+2S_y\right) ^2-\left( m_2+2T_y\right) ^2\right]
\label{p2.9}
\end{equation}
For convenience, we denoted $m_1=\pi \left( n_2-n_1\right) /a_x$, $m_2=\pi
\left( n_3-n_1\right) /a_x$, where $n_i$ is the summation index
corresponding to the order parameter $\Delta \left( \overrightarrow{r}%
_i\right) $ in Eq.(\ref{p2.6a}). Note that only a 2D sum remains in Eq.(\ref
{p2.9}) from the original 4D sum; the free double sum is equal to the number
of vortices-$N$, while the summation in Eq.(\ref{p2.9}) is invariant under
the vortex number shift transformation $n_i\rightarrow n_i+n_0$. Note also
that, the combination of gauge factors from the order parameters and Green's
function causes integral over center mass coordinate $\overrightarrow{R}$ to
vanish if $n_1-n_2+n_3-n_4\neq 0$.

It is clear from Eq.(\ref{p2.9}) that the lattice sum in ${\cal D}\left( 
\overrightarrow{S},\overrightarrow{T}\right) $ is dominated by the lattice
point $m_1\simeq -2S_y$, $m_2\simeq -2T_y$. Taking only this term in the sum
into account we obtain that 
\begin{equation}
{\cal D}\left( \overrightarrow{S},\overrightarrow{T}\right) \propto
Ne^{-4i\left( 2\gamma S_yT_y+S_yT_x+S_xT_y\right) }  \label{p2.10}
\end{equation}
The lattice function ${\cal D}\left( \overrightarrow{S},\overrightarrow{T}%
\right) $ being multiplied by the kernel $\widetilde{K}_{4,\nu }\left(
\left\{ \rho _l\right\} \right) $ determines the free energy distribution in
the space of relative electron coordinates.

\section{Asymptotic local approximation}

In this section we analyze in detail the relative importance of all
different spatial regions contributing to the multiple integral , Eq.(\ref
{p2.8}) , which determines the quartic term in the asymptotic limit $\sqrt{%
n_F}\gg 1$.

Every four electron configuration in this integral is defined by the three
vectors $\left( \overrightarrow{Q},\overrightarrow{S},\overrightarrow{T}%
\right) $. Since the kernel $\widetilde{K}_{4,\nu }\left( \overrightarrow{Q},%
\overrightarrow{S},\overrightarrow{T}\right) $ is a bounded function at $%
\left| \overrightarrow{Q}\right| >>1$, we may conclude from Eq.(\ref{p2.8})
that the main contribution to the free energy comes from the the region $%
\left| \overrightarrow{Q}\right| \leq 1$ , and so assume in what follows
that $\left| \overrightarrow{Q}\right| \leq 1$. To estimate the integrals
over $\overrightarrow{S}$ and $\overrightarrow{T}$ we separate the entire
domain of integrations into three regions according to the behavior of the
Green's function (Eq.(\ref{p2.2a})) and the function ${\cal D}\left( 
\overrightarrow{S}, \overrightarrow{T}\right)$.

1. The turning point region $\rho _i\simeq 2r_F$, for which in the
asymptotic limit $\sqrt{n_F}\gg 1$, 
\[
\rho _{1,3}=2\left| \overrightarrow{Q}\pm \overrightarrow{S}\right| \simeq
2S\simeq 2r_F=2\sqrt{2n_F}\gg 1 
\]
\begin{equation}
\rho _{2,4}=2\left| \overrightarrow{Q}\mp \overrightarrow{T}\right| \simeq
2T\simeq 2r_F=2\sqrt{2n_F}\gg 1,  \label{p3.1}
\end{equation}
The size of this region is of the order $\Delta \rho _i=\Delta S=\Delta
T\sim 1/r_F^{1/3}$. It is characterized by a smooth behavior of all Green's
functions $\widetilde{G}_0\left( \rho _i\right) \sim \widetilde{G}_0\left(
2r_F\right) \sim 1/n_F^{1/3}$ and kernel $\widetilde{K}_{4,\nu }\sim \left[ 
\widetilde{G}_0\left( 2r_F\right) \right] ^4\sim 1/n_F^{4/3}$.

2. In the intermediate region $1\ll \rho _i<2r_F$ variables $S$ and $T$ are
still large: $S\gg 1$, $T\gg 1$ but essentially less then $r_F$. Here the
Green's function is approximated by the expression Eq.(\ref{p2.2b}). To
simplify the considerations we note that the oscillating phase factor in Eq.(%
\ref{p2.2b}) can be written as 
\begin{equation}
n_F\left( \phi +\sin \phi \right) =\sqrt{2n_F}\xi \left( \rho /2r_F\right)
\rho  \label{p3.2}
\end{equation}
where $\pi /4\leq \xi \left( x\right) \leq 1$ at $0\leq x\leq 1$. For our
purpose we may take $\xi \left( x\right) =1$. Substituting $\rho
_{1,3}\simeq 2S$, $\rho _{2,4}\simeq 2T$, the kernel $\widetilde{K}_{4,\nu }$
in this region can be transformed to 
\begin{equation}
\widetilde{K}_{4,\nu }\left( \overrightarrow{Q},\overrightarrow{S},%
\overrightarrow{T}\right) \propto \frac{\exp \left[ i4\sqrt{2n_F}\left(
S-T\right) \right] }{ST\left( r_F^2-S^2\right) ^{1/2}\left( r_F^2-T^2\right)
^{1/2}}  \label{p3.3}
\end{equation}

3. In the third region all variables $\overrightarrow{Q},\overrightarrow{S},%
\overrightarrow{T}$ are of the order of the magnetic length: $Q,S,T\sim 1$.
Here ${\cal D}\left( \overrightarrow{S},\overrightarrow{T} \right) $ becomes
a smooth function and the strong oscillations of the integrand in Eq.(\ref
{p2.8}) arise only from $\widetilde{K}_{4,\nu }$. Since $\xi \left( x\right)
\rightarrow 1$ at $x\rightarrow 0$ the exact short distance asymptotic is 
\begin{equation}
\widetilde{K}_{4,\nu }\left( \left\{ \rho _i\right\} \right) \propto \frac 1{%
n_F\left( \rho _1\rho _2\rho _3\rho _4\right) ^{1/2}}\exp \left[ i\sqrt{2n_F}%
\left( \rho _1-\rho _2+\rho _3-\rho _4\right) \right]  \label{p3.4}
\end{equation}

Let us first consider the case when both variables $\overrightarrow{S},%
\overrightarrow{T}$ belong to the first region (i.e. region(1,1)): Replacing 
$\widetilde{K}_{4,\nu }\left( \overrightarrow{Q},\overrightarrow{S} ,%
\overrightarrow{T}\right) $ by a constant of the order $\sim 1/n_F^{4/3}$ we
can estimate the free energy ,Eq.(\ref{p2.8}), for a given Matsubara
frequency in polar coordinate system $\overrightarrow{S}\equiv \left(
S,\alpha _s\right) $, $\overrightarrow{T}=\left( T,\alpha _t\right) $ as 
\begin{equation}
F_{s,\nu }^{\left( 4\right) }\propto \frac 1{n_F^{4/3}}\int\limits_{r_F-%
\Delta \rho }^{r_F}SdSTdT\int\limits_0^{2\pi }d\alpha _sd\alpha
_te^{-4iST\phi \left( \alpha _s,\alpha _t\right) }  \label{p3.5}
\end{equation}
where $\phi \left( \alpha _s,\alpha _t\right) =\gamma \cos \left( \alpha
_s-\alpha _t\right) +\sin \left( \alpha _s+\alpha _t\right) -\gamma \cos
\left( \alpha _s+\alpha _t\right) $. After simple integration over $\alpha
_s $ and $\alpha _t$ by the stationary phase method, which is justified by
the very large values of $ST$, Eq.(\ref{p3.5}) transforms to 
\begin{equation}
F_{s,\nu }^{\left( 4\right) }\propto \frac 1{n_F^{4/3}}\int\limits_{r_F-%
\Delta \rho }^{r_F}dSdTe^{-4iST\phi _s^{\pm }}\sim \frac 1{n_F^{7/3}}
\label{p3.5a}
\end{equation}
where $\phi _s^{\pm }=\sqrt{1+\gamma ^2}\pm \gamma $ are the values of $\phi
\left( \alpha _s,\alpha _t\right) $ at the stationary points.

Let us next consider the case when both $S$ and $T$ belong to the
intermediate region (i.e. region (2,2)): Here $\widetilde{K}_{4,\nu } \left(
\left\{ \rho _i\right\} \right) $, defined by Eq.(\ref{p3.3}) like ${\cal D}%
\left( \overrightarrow{S},\overrightarrow{T}\right) $, is a sharply
oscillating function with frequency of order of $\sqrt{n_F}$. Substituting (%
\ref{p3.3}),(\ref{p2.10}) into (\ref{p2.8}) we obtain for $1\ll S,T<r_F$%
\begin{equation}
F_{s,\nu }^{\left( 4\right) }\propto \int \frac{d^2Sd^2T\exp \left[ 4i\left( 
\sqrt{2n_F}\left( S-T\right) -2\gamma S_yT_y-S_yT_x-S_xT_y\right) \right] }{%
ST\left( r_F^2-S^2\right) ^{1/2}\left( r_F^2-T^2\right) ^{1/2}}  \label{p3.6}
\end{equation}
The integrals over angle variables are identical to previously considered
ones and give a factor of the order $1/ST$. Thus free energy (\ref{p3.6})
reduces to $F_{s,\nu}^{\left( 4\right) }\propto \int \frac{dSdT\exp \left[
4i\left( \sqrt{2n_F}\left( S-T\right) \pm ST\phi _s^{\pm }\right) \right] }{%
ST\left( r_F^2-S^2\right) ^{1/2}\left( r_F^2-T^2\right) ^{1/2}}$. Taking for
the smooth preexponential factor its value at a point $S\sim \sqrt{n_F} $, $%
T\sim \sqrt{n_F}$ and noting that there is no stationary point for the
integrals over $S$ and $T$, we get the following estimate for free energy in
the intermediate region: $F_{s,\nu}^{\left( 4\right) }\sim 1/n_F^3$.

It is clear that if one of variables, for example $S$, is from the turning
point region and the other, $T$, from the intermediate region (i.e. region
(1,2)), the free energy $F_{s,\nu}^{\left( 4\right) }$ will be proportional
to a factor of the order $\sim 1/n_F^{8/3}$. This result follows from the
fact that the product of two of the Green's function are proportional to $%
\sim 1/n_F^{2/3}$, the product of the other two is proportional to $\sim
1/n_F$ and the integration over $S$ and $T$ yields the factor $\sim 1/n_F$.
The integration over the angles $\alpha _s$ and $\alpha _t$ produces the
factor $\sim 1/ST$ which cancels the same factor in the numerator of the
integrand (\ref{p2.8}).

Let us consider now the four-electron configuration when one of the
variables, $S$ or $T$, e.g. $T$ , is of the order of the magnetic length $%
T\sim 1$ (i.e. region (3)). In contrast to previous cases, $S,T\gg 1$, where
the oscillations with $\overrightarrow{Q}$ can be neglected, in this case an
additional small factor arises from the integration over $\overrightarrow{Q}$%
. It is partially cancelled by the large factor arising from the integration
over $\overrightarrow{S}$ if it is from the turning point region. Assuming
that $S\sim r_F$ and $T\sim 1$ (i.e. region (1,3)), the estimate of the free
energy (\ref{p2.8}) is given by the formula 
\begin{eqnarray}
F_{s,\nu }^{\left( 4\right) } &\propto &\frac 1{n_F^{7/6}}\int
d^2Qe^{-4Q^2}\int d^2Sd^2T \\
&&\exp \left[ -4i\left( \frac 12\sqrt{2n_F}\left( \left| \overrightarrow{Q}-%
\overrightarrow{T}\right| +\left| \overrightarrow{Q}+\overrightarrow{T}%
\right| \right) +2\gamma S_yT_y+S_yT_x+S_xT_y\right) \right]  \label{p3.7}
\end{eqnarray}
The integral over $\overrightarrow{Q}$ can be divided into two regions: $%
Q\leq T$ and $Q\geq T$. The main contribution comes from $Q\leq T$, where
the above exponential factor does not depend on the component $Q^{\parallel
} $, parallel to $\overrightarrow{T}$ . For the small transverse component $%
Q^{\perp }$ we get 
\begin{equation}
\left| \overrightarrow{Q}-\overrightarrow{T}\right| +\left| \overrightarrow{Q%
}+\overrightarrow{T}\right| \simeq 2T+ \left( \frac 1{\rho _2}+\frac 1{\rho
_4}\right) \left( Q^{\perp }\right) ^2  \label{p3.7a}
\end{equation}
Thus the integral over $Q^{\perp }$ is proportional to $\sim 1/n_F^{1/4}$
and the integral over $Q^{\parallel }$ gives $\int_0^TdQ^{\parallel }\sim T$%
. Performing the integration over angles $\alpha _s$ and $\alpha _t$ gives,
as previously, $1/ST$ and taking into account that integral over $S$ from a
smooth function can be estimated by the area of the integration region: $%
S\Delta S\sim r_F\Delta \rho \sim n_F^{1/3}$, we get for $Q\leq T$ that 
\begin{equation}
F_{s,\nu }^{\left( 4\right) }\sim \frac 1{n_F^{19/12}}\int TdT\exp \left( -4i%
\sqrt{n_F}T\left( 1\pm \phi _s^{\pm }\right) \right) \sim \frac 1{n_F^{31/12}%
}  \label{p3.7b}
\end{equation}

In the second region ,$Q\geq T$, the integral over $\overrightarrow{Q}$
leads to a contribution smaller by the factor $n_{F}^{1/4}$ than (\ref{p3.7b}%
). This is because the phase (\ref{p3.7a}) does not depend on $T^{\parallel
} $ and we can neglect the quadratic in $T^{\perp }$ term in comparison with
the linear one in Eq.(\ref{p3.7}).Thus the integral over $\overrightarrow{Q}$
can be estimated here as $\sim 1/n_F$.

If $\overrightarrow{S}$ belongs to the intermediate region (i.e. considering
the region (2,3)), the contribution to the free energy will be even smaller
because of the Green's function oscillations and smaller preexponential
factor.

Finally, in the short distance region $\rho _i\leq 1$ the smooth function $%
{\cal D} \left( \overrightarrow{S},\overrightarrow{T}\right) $ is of the
order one and we should calculate the integrals in (\ref{p2.8}) from the
kernel $\widetilde{K}_{4,\nu }\left( \left\{ \rho _i\right\} \right) $ Eq.(%
\ref{p3.4}). The phase factor of $\widetilde{K}_{4,\nu }$ which is
proportional to 
\begin{eqnarray}
\Phi \left( \overrightarrow{Q},\overrightarrow{S},\overrightarrow{T}\right)
&=&\rho _1-\rho _2+\rho _3-\rho _4=  \nonumber \\
&=&2\left\{ \left| \overrightarrow{Q}+\overrightarrow{S}\right| -\left| 
\overrightarrow{Q}-\overrightarrow{T}\right| +\left| \overrightarrow{Q}-%
\overrightarrow{S}\right| -\left| \overrightarrow{Q}+\overrightarrow{T}%
\right| \right\}  \label{p3.8}
\end{eqnarray}
at fixed $Q$ has a set of stationary points $S\leq Q$, $T\leq Q$ for
collinear vectors $\overrightarrow{Q}$,$\overrightarrow{S}$,$\overrightarrow{%
T}$. Going back to $\overrightarrow{\rho }_i$ variables we conclude that
this configuration is equivalent to the propagation of ''odd'' particles in
a single direction $\overrightarrow{n}$ and ''even'' particles in the
opposite direction: 
\begin{equation}
\overrightarrow{\rho }_1=\rho _1\overrightarrow{n},\overrightarrow{\rho }%
_2=-\rho _2\overrightarrow{n},\overrightarrow{\rho }_3=\rho _3%
\overrightarrow{n},\overrightarrow{\rho }_4=-\rho _4\overrightarrow{n},
\label{p3.9}
\end{equation}
where $\overrightarrow{n}$ is an arbitrary unit vector. Note that the
special role of the configuration (\ref{p3.9}) follows from the fact that
for this configuration $\sum \overrightarrow{\rho }_i=\Phi \left( \rho
_i\right) \overrightarrow{n}\equiv 0$ in the integration region.

It should be emphasized that in the resulting coherent configuration the
correlation among all four particles is essential. The phase factor, Eq.(\ref
{p3.8}), and stationary point equations can not be factorized.

Expanding $\Phi\left( \overrightarrow{Q},\overrightarrow{S}, \overrightarrow{%
T}\right)$ in the coordinates $S^{\perp }$ and $T^{\perp }$ perpendicular to 
$\overrightarrow{n}$ , and noting that from the definition of $%
\overrightarrow{Q}$ follows $\overrightarrow{n}=-\overrightarrow{Q}/Q$, we
can reduce (\ref{p3.8}) to 
\begin{equation}
\Phi \left( \overrightarrow{Q},\overrightarrow{S},\overrightarrow{T}\right)
=2\left[ \left( \frac 1{\rho _1}+\frac 1{\rho _3}\right) \left( S^{\perp
}\right) ^2+\left( \frac 1{\rho _2}+\frac 1{\rho _4}\right) \left( T^{\perp
}\right) ^2\right]  \label{p3.10}
\end{equation}
where $\rho _1+\rho _3=\rho _2+\rho _4=4Q$. Now performing the integration
of the kernel $\widetilde{K}_{4,\nu }\left( \left\{ \rho _i\right\} \right)$
, given in Eq.(\ref{p3.4}), over $S^{\perp }$ and $T^{\perp }$ we obtain a
factor $\sim \frac 1{\sqrt{n_F}}\frac{\left( \rho _1\rho _2\rho_3\rho
_4\right) ^{1/2}}Q$, which gives rise to the $\nu$ dependent free energy $%
F_{s,\nu}^{\left(4\right) }\sim 1/n_F^{3/2}$.

This result completes our analysis: it implies that in the asymptotic limit ,%
$\sqrt{n_{F}}\gg 1$, the dominant contributions to the quartic term of the
free energy originate in the short distance region only. Thus, using the
short distance approximation for $\widetilde{K}_{4,\nu }$, taking into
account its dependence on $\omega _\nu $, and restoring the exact form of $%
{\cal D}\left( \overrightarrow{S}, \overrightarrow{T}\right) $ ,Eq.(\ref
{p2.9}), the quartic term of the SC free energy can be written in the form:

\begin{eqnarray}
F_s^{\left( 4\right) } &\propto &\frac 1{n_F^{3/2}}\sum\limits_\nu q_\nu
^2\int dQe^{-4Q^2-4\alpha _\nu Q}\int_{-Q}^QdSdTd\theta  \label{p3.11} \\
&&\ \exp \left[ -4\left( S^2+T^2\right) \sin ^2\theta +4iST\sin \left(
2\theta \right) \right] \sum\limits_{m_1,m_2}\exp \left[ -m_1^2-m_2^2\right.
\nonumber \\
&&\ \left. -2i\gamma m_1m_2+4i\left( m_1T+m_2S\right) \cos \theta -\left(
m_1S+m_2T\right) \sin \theta \right]  \nonumber
\end{eqnarray}
where the angle $\theta$ describes the direction of the unit vector $%
\overrightarrow{n}$ , $2q_\nu =J\left( -\omega _\nu \right) J\left( \omega
_\nu \right) $, and $\alpha_{\nu}=2(2\nu +1)a_{H}/\zeta$, with $\zeta=\hbar
v_{F}/(\pi k_{B}T)$.

This free energy is determined by collinear, essentially four particles
configurations with the size of the order of magnetic length. The resulting
expression,Eq.(\ref{p3.11}), agrees with the quartic term $F_s^{(4)}$
derived previously by Maniv et al.\cite{man92}. Our present considerations
justify the used approximation and clarify the geometry of coherent
configurations.

It should be emphasized that the random vortex lattice approximation, used
in \cite{ste92}, gives rise to a markedly different result, namely $%
F_{s,\nu}^{\left( 4\right) }\sim 1/n_F$. The reason for the disagreement is
due to the averaging over random vortex lattice \cite{ste92} which leads to
factorization of the multiple products of pair potentials into products of
pair correlation functions only.

For example, the quartic term in this approximation becomes 
\[
\left\langle \Delta \left( \overrightarrow{r}_1\right) \Delta ^{\star
}\left( \overrightarrow{r}_2\right) \Delta \left( \overrightarrow{r}%
_3\right) \Delta ^{\star }\left( \overrightarrow{r}_4\right) \right\rangle
\propto \left\langle \Delta \left( \overrightarrow{r}_1\right) \Delta
^{\star }\left( \overrightarrow{r}_2\right) \right\rangle \left\langle
\Delta \left( \overrightarrow{r}_3\right) \Delta ^{\star }\left( 
\overrightarrow{r}_4\right) \right\rangle 
\]

\[
\propto \exp \left[ -\frac{1}{2}\left( \rho _1^2+\rho _3^2\right) \right]
\exp \left[ -i\left( \zeta \left( 2,1\right) +\zeta \left( 4,3\right)
\right) \right] 
\]

where $\zeta \left( 2,1\right) =\left( x_2+x_1\right) \left( y_2-y_1\right) $
is the (Landau) gauge factor of $G_{0}(\overrightarrow{r}_{1}, 
\overrightarrow{r}_{2};\omega_{\nu})$, and $\left\langle ....\right\rangle $
stands for averaging over vortex distributions. The corresponding free
energy is given by 
\[
F_{s,\nu}^{\left( 4\right) }\propto \int \prod d^2r_i\widetilde{K}_{4,\nu }
\left(\left\{ \rho _i\right\} \right) \exp \left[ -\frac{1}{2}\left( \rho
_1^2+\rho _3^2\right) \right] \times 
\]
\begin{equation}
\exp \left[ \frac{i}{2}\left[ \left( x_4-x_2\right) \left( y_3-y_1\right)
-\left( x_3-x_1\right) \left( y_4-y_2\right) \right] \right]  \label{p2.15}
\end{equation}
In the important region of integration $\rho _1,\rho _3\leq 1$, the
constraint $\sum_{l=1}^{4}\overrightarrow\rho_{l}=0$, implies that $%
\overrightarrow\rho_{2}\approx -\overrightarrow\rho_{4}$, or alternatively $%
\overrightarrow{r}_1\sim \overrightarrow{r}_2$ and $\overrightarrow{r}_3\sim 
\overrightarrow{r}_4$. Thus the imaginary exponent in Eq.(\ref{p2.15}) is
always of the order unity or smaller in the important region of
integrations, and so the total gauge factor is a smooth function.
Consequently, only two distances in the kernel $\widetilde{K}_{4,\nu }$ are
restricted to the size of the order of magnetic length, allowing two others
to be arbitrary. Thus, in contrast to our result, where all long range
configurations of electron pairs in the ordered vortex lattice interfere
destructively (i.e. appear as incoherent 'scattering' channels) , the
smoothing of rapidly oscillating gauge factors in the MS theory introduces a
huge 'incoherent' contribution to the SC free energy.

Let us estimate now the quartic term within this approximation. Substituting
for the Green's functions $\widetilde{G}_0\left( \rho
_1,-\omega_{\nu}\right) $ and $\widetilde{G}_0\left( \rho
_3,-\omega_{\nu}\right)$ their approximants in the short distance region,
and omiting the smooth gauge factor, we get for the free energy $%
F_{s,\nu}^{\left(4\right) }$, after integrating over the center mass
coordinates:

\begin{eqnarray}
F_{s,\nu }^{\left( 4\right) } &\propto &\frac 1{n_F^{1/2}}\int \frac{\prod
d^2\rho _i}{\left( \rho _1\rho _3\right) ^{1/2}}\delta \left( \sum_{l=1}^4%
\overrightarrow{\rho }_l\right) \widetilde{G}_0\left( \rho _2,\omega _\nu
\right) \widetilde{G}_0\left( \rho _4,\omega _\nu \right)  \nonumber \\
&&\ \exp \left( i\sqrt{2n_F}\left( \rho _1+\rho _3\right) \right) \exp
\left[ -\frac 12\left( \rho _1^2+\rho _3^2\right) \right]  \label{p2.15a}
\end{eqnarray}

The main contribution to Eq.(\ref{p2.15a}) arises from the region $\rho
_2\simeq 2r_F$, $\rho _4\simeq 2r_F$. Allowing $\overrightarrow{\rho} _2$, $%
\overrightarrow{\rho} _4$ to vary independently within the turning point
region, $\overrightarrow{\rho }_1$, $\overrightarrow{\rho }_3$ are not
independent variables; taking $\overrightarrow{\rho }_1$ as the third
independent variable of integration , and noting that $\overrightarrow{\rho }%
_1\simeq -\overrightarrow{\rho }_3$, we have 
\begin{equation}
F_{s,\nu}^{(4)}\sim \frac{1}{n_F^{1/2}}\int d^2\rho_1 \frac{e^{2i\sqrt{2n_F}%
\rho_1}}{\rho_1}\left[\int d^2\rho \tilde{G}_0(\rho,\omega_{\nu})\right]^2
\end{equation}

Now, since $\int \widetilde{G}_0\left( \rho \right) d^2\rho \sim 1$, and the
integration over $\overrightarrow{\rho }_1$ yields the factor $\sim 1/\sqrt{%
2n_F}$, the resulting $n_F$ dependence is $1/n_F$, in agreement with \cite
{ste92}.

\section{Self Consistent Order Parameter}

The local approximation, verified in the previous section, becomes very
transparent if we rewrite the free energy ,Eq.(\ref{p3.11}), as a functional
of the order parameter $\Delta(\overrightarrow{r})$ . After some
straightforward, but combersome calculations one can show that Eq.(\ref
{p3.11}) is equivalent to 
\begin{eqnarray}
F_s^{\left( 4\right) } &\propto &\frac 1{n_F^{3/2}}\int d^2R\int d\theta
dQf\left( Q\right) e^{-4Q^2}\int\limits_{-Q}^QdS\int\limits_{-Q}^QdT\Delta
\left( \overrightarrow{R}+\left( S+T\right) \overrightarrow{n}\right)
\label{p3.12} \\
&&\Delta ^{\star }\left( \overrightarrow{R}+\left( S-T\right) 
\overrightarrow{n}\right) \Delta \left( \overrightarrow{R}-\left( S+T\right) 
\overrightarrow{n}\right) \Delta ^{\star }\left( \overrightarrow{R}-\left(
S-T\right) \overrightarrow{n}\right)  \nonumber
\end{eqnarray}
where $f\left( Q\right) =\sum\limits_\nu q_\nu ^2e^{-4\alpha _\nu Q}$, and $%
\Delta \left( \overrightarrow{R}\right) $ is defined by Eq.(\ref{p2.6a}).
Since $\left| S\right| ,\left| T\right| \leq Q\leq 1$, the expression (\ref
{p3.12}) can be considered as averaging of the four order parameter product
over a region with radius of the order of the magnetic length. The
additional averaging over the direction of $\overrightarrow{n}$ in (\ref
{p3.11}) leads to a completely local expression plus a nonlocal correction,
i.e.:

\begin{equation}
F_s^{\left( 4\right) }=B\int d^2R\left| \Delta \left( \overrightarrow{R}
\right) \right| ^4+F_{s,nloc}^{\left( 4\right) }  \label{p3.13}
\end{equation}
where $B \propto \frac 1{n_F^{3/2}}\int dQf\left( Q\right)
e^{-4Q^2}\int\limits_{-Q}^QdS\int\limits_{-Q}^QdTe^{-2\left( S^2+T^2\right)
} $. The nonlocal correction , $F_{s,nloc}^{(4)}$, is numerically small
since it arises from high (i.e fourth and higher) order terms in the
'cumulant' expansion of the exponential in Eq.(\ref{p3.11}) (see Ref.\cite
{man92}).

This result is of fundamental importance since it shows that the well known,
fully local form of the Ginzburg-Landau free energy functional in the low
field regime near $T_{c}(H=0)$ is basically valid also in the opposite ,
high magnetic field regime near $H_{c2}(T=0)$. This locality is closely
related to the coherence effect discussed above. For example, in the random
lattice approximation, discussed in the previous section, the dominant
contribution to $F_{s}^{(4)}$ is extremely nonlocal.

Neglecting the small nonlocal correction, the total SC free energy, up to
fourth order in $\Delta_{0}$, can be turned into the following one parameter
variational form \cite{man92}: 
\begin{equation}
f_{s}\equiv\frac{F_{s}}{N\pi a_{H}^2}={\cal D}_{2D}\left[ -\tilde{\alpha}
\Delta_{0}^2 +\frac{\tilde{B}}{(\pi k_B T_c)^2} \Delta_{0}^4\right]
\label{var}
\end{equation}

where ${\cal D}_{2D}=m_c/2\pi\hbar^2$ (i.e. the 2D single electron density
of states), 
\begin{equation}
\tilde{\alpha}=2\frac{a_H}{\zeta}\sum_{\nu=0}^{\nu_D }Re(q_{\nu})
\gamma_{\nu}-1/g  \label{quadr}
\end{equation}
with $\gamma_{\nu}=\int_0^{\infty}d\rho e^{-\alpha_{\nu}\rho -\frac{ 1}{2}
\rho^2}$, $g=V{\cal D}_{2D}$ , and $\nu_D\equiv(T_D/2T-1)$, where $T_D$ is
the Debye temperature.

The coefficient ,$\tilde{B}$, of the quartic term can be readily obtained
from Eq.(\ref{p3.13}) (after replacing $Q$ with $\rho/2$): 
\begin{equation}
\tilde{B}=\beta_A\frac{a_H}{\zeta}(\frac{a_H}{\xi_0})^2\sum_{\nu}^ {\nu_D
}Re({q_{\nu}^2})\delta_{\nu}
\end{equation}
with $\delta_{\nu}\equiv 2\pi\int_0^{\infty}d\rho e^{-2\alpha_{\nu}\rho -
\rho^2}erf^{2}(\rho/\sqrt{2})$, $\xi_0\equiv\hbar v_F/\pi k_B T_c$, and $%
\beta_A$ is the geometrical factor of the Abrikosov lattice \cite{man92}.

The key parameters , which control the crossover from the low field to the
high field regime, are $a_H/\zeta$, and $X=2\pi^2 k_B T/\hbar\omega_c$; they
are connected by: 
\begin{equation}
X=2\pi(2n_F)^{1/2}(\frac{a_H}{\zeta})
\end{equation}
which means that in the asymptotic limit considered here, our 'high'
temperatue regime, $X\sim 1$, of the quantum magnetic oscillations domain is
still in the low temperature regime of the SC-normal phase boundary, since $%
\frac{a_H}{\zeta}\sim 1/\sqrt{n_F}\ll 1$

In this case $q_{\nu}\approx 2$ for all $\nu$, and the coefficient, $\tilde{%
\alpha}$, of the quadratic term, can be calculated from Eq.(\ref{quadr}) by
dividing the sum over the Matzubara frequencies ,$\nu$, into two regions:
(1) $\alpha_{\nu}\ll 1$, namely $\nu\ll \nu_{max} \equiv \frac{\zeta}{a_H}/2%
\sqrt{2}$, and (2) $\nu \ge \nu_{max}$. The contribution from the first
region is $\sqrt{\pi}\int_0^{1}e^{x^2}(1-erf(x))dx\approx 1.147$, while the
sum in the second region leads to the familiar logarithmic expression $%
\sum_{\nu=\nu_{max}}^{\nu_D }1/(\nu +1/2) \approx\ln{(\sqrt{2}\frac{T_D}{T}%
\frac{a_H}{\zeta})}$, provided that the Debye cut-off temperatue $T_D\equiv
(2\nu_{D}+1)T$ is much larger than $(2\nu_{max}+1)T$. The last condition may
be rewritten in a more transparent form, i.e. $(k_B T_D/\hbar\omega_c)^2\gg
n_F/2\pi^3$.

Combining the contributions from the two regions we find 
\begin{equation}
\tilde{\alpha}\approx \ln\left[\frac{a_H}{\sqrt{2}\xi(0)}\right]
\end{equation}
where $\xi(0)\equiv .18\hbar v_F/k_B T_c \approx .56\xi_0$.

Now consider the coefficient $\tilde{B}$, of the quartic term. Again, we
divide the Matzubara sum into the same two regions: In the first, where $\nu
\ll \nu_{max}$, each term,$\delta_{\nu}$, is independent of $\nu$ so that $%
\sum_{\nu=0}^{\nu_{max}}\delta_{\nu}\approx 4\nu_{max}\int_{0}^{\infty}
d\rho e^{-\rho^2}\rho^2 d\rho\approx .63(\frac{\zeta}{a_H})$, whereas the
second region yields $\sum_{\nu=\nu_{max}}^{\infty} \left[\frac{1}{2(2\nu
+1)a_H/\zeta}\right]^3\approx \frac{1}{16} \zeta/a_H$. Combining these
results we find that the sum over Matzubara frequencies changes
significantly the $n_F$ dependence of the quartic term with respect to the
indevidual $F_{s,\nu}^{(4)}$ terms, since $\sum_{\nu}\delta _{\nu}\approx
.69(\frac{\zeta}{a_H})\sim n_{F}^{1/2}$.

We thus find that $\tilde{B}/(\pi k_B T_c)^2 \approx 1.38/E_F\hbar\omega_c$,
so that 
\begin{equation}
f_s\approx \frac{\hbar\omega_c}{2\pi a_H^2}\left[-\tilde{\Delta}_{0} ^{2}\ln{%
\left(\frac{a_H}{\sqrt{2}\xi(0)}\right)}+\frac{1.38}{n_F} \tilde{\Delta}%
_{0}^{4}\right]  \label{fs}
\end{equation}
It should be emphasized, here again, that the $n_F$ dependence of both the
quartic and the quadratic terms in $f_s$ above differs by the large factor $%
\sqrt{n_F}$ from the indevidual terms $F_{s,\nu}^{(4)}$,$F_{s,\nu}^{(2)}$
because of the sum over the Matzubara frequencies.

Using Expression (\ref{fs}), the self consistent mean field order parameter
is given by 
\begin{equation}
\tilde{\Delta}_0^{2}=.36n_F \ln{\left(\frac{a_H}{\sqrt{2}\xi(0)} \right)}
\label{delta0}
\end{equation}

This expression is identical to the well known high field limit of the
Gorkov-Ginzburg-Landau SC order parameter (\cite{wass96}). Indeed, at
magnetic fields $H$ near $H_{c2}(0)=\phi_0/2\pi\xi(0)^2$, $\phi_0 =ch/2e$,
where $a_H\approx\sqrt{2}\xi(0)$, we have: 
\[
E_F\hbar\omega_c\approx\frac{(\pi k_B T_c)^2}{2(a_H/\xi_0)^2} \approx
.78(\pi k_B T_c)^2 
\]
, so that Eq.(\ref{delta0}) reduces to the well known result 
\begin{equation}
\Delta_0\approx 1.7k_B T_c\left[\ln{\left(\frac{a_H}{\sqrt{2}\xi(0)}\right)}
\right]^{1/2}\approx 1.7k_B T_c\left[1-H/H_{c2}(0)\right]^{1/2}
\end{equation}

Interestingly, the $n_F$ dependence of the self consistent $\Delta _0$
obtained in Eq.(\ref{delta0}) for $H\approx H_{c2}$ determines a small
parameter: 
\[
x\equiv \frac{\tilde \Delta _0^2}{n_F}\approx .36\left[ 1-H/H_{c2}(0)\right] 
\]
which is seen to be the expansion parameter in the perturbation theory
leading to Eq.(\ref{fs}). This observation will be further discussed in the
next section.

Note that in deriving the above expressions for the self consistent order
parameter we have neglected the oscillatory compomemts of the SC free
energy, which should add an oscillatory contribution to the order parameter 
\cite{man92,bruun97}. In the 'high temperature' limit considered, this
oscillatoray term is much smaller than the nonoscillatory one, except for a
very narrow region near $H_{c_2}$ \cite{rom95}.

Let us consider now the magnetization oscillations; the dominant
contribution to the superconducting part can be obtained by differentiating
the density of states factors $q_{\nu}$ in the free energy (\ref{var}) with
respect to magnetic field, namely: 
\begin{equation}
M_{s,osc}\propto -\sum_{\nu}\frac{\partial f_s}{\partial q_{\nu}} \frac{%
\partial q_{\nu}}{\partial H}
\end{equation}

Explicitly we have: 
\begin{equation}
M_{s,osc}\approx 2{\cal D}_{2D}\frac{a_H}{\zeta}\Delta_0^2
\sum_{\nu=0}^{\nu_D -1}\left[\gamma_{\nu}-(\frac{\Delta_0} {\pi k_B T_c})^2(%
\frac{a_H}{\zeta})\delta_{\nu}q_{\nu}\right] \frac{\partial q_{\nu}}{%
\partial H}
\end{equation}

For $X\ge 1$, $\frac{\partial q_{\nu}}{\partial H}\approx -(8\pi n_F/H)\sin{%
(2\pi n_F)}e^{-(2\nu +1)X}$, so that the sum over $\nu$ is limited by the
thermal damping factor to the first few terms only. This contrasts the
nonoscillatory magnetization, which picks up contributions from many
Matzubara frequencies.

Thus the first harmonic of the oscillatory magnetization, $M_{osc}$, just
below $H_{c2}$ can be written as \cite{man94} 
\begin{equation}
\widetilde{M}_{osc}\equiv \frac{\phi _0}{E_F}M_{osc}\approx \widetilde{M}%
_{n,osc}\left[ 1-\frac{\pi ^{3/2}\widetilde{\Delta }_0^2}{n_F^{1/2}}+\frac{%
\sqrt{2}\pi ^{3/2}\beta _A\widetilde{\Delta }_0^4}{n_F^{3/2}}\right]
\label{p2.16}
\end{equation}
where $\beta _A\approx 1.16$ for a triangular lattice , and $\widetilde{M}%
_{n,osc}\equiv \frac{X}{e^X} \sin \left( 2\pi n_F\right)$ is the normal
electrons oscillatory magnetization \cite{sho84}.

In the expansion (\ref{p2.16}) there are two scales of order parameter $%
\Delta _0$. Near $H_{c2}$, where $\Delta _0^2\leq \left( \hbar \omega
_c\right) ^{3/2}E_F^{1/2}$ (i.e. $\tilde{\Delta}_0^2\leq n_F^{1/2}$, which
means that $\ln{[\frac{a_H}{\sqrt{2}\xi(0)}]}\sim 1/n_F^{1/2}$ ), the
attenuation of the magnetization oscillations amplitude occurs as the result
of the electron pairing. Here the contribution of the many electron coherent
configurations are negligeable. Far from $H_{c2}$, where $\Delta _0^2\sim
\hbar \omega _cE_F$ (i.e. $\tilde{\Delta}_0^2\sim n_F$ so that $\ln{[\frac{%
a_H} {\sqrt{2}\xi(0)}]}\sim 1$ ) , the quadratic and the quartic terms in
the free energy (and magnetization) are comparable. It can be shown \cite
{zhu} that the higher order terms in this expansion are determined by the
parameter $\Delta _0^2/\hbar \omega _cE_F=\frac{\tilde{\Delta}_0^{2}}{n_F}$.
In the region where this parameter is of the order unity or larger the SC
state is a highly correlated many electron-pair configuration , which is
quite different from the condensate of electron pairs, dominating the SC
free energy just below $H_{c2}$.

\section{Conclusion}

The results of the last two sections enable us now to critically discuss the
various theoretical approaches to the problem of the intrinsic attenuation
of the dHvA oscillations in the vortex state, and the relevance of our model
to real experiemts.

It is, first of all, clear that the assumption of disordered vortex lattice,
and the consequent averaging over the random pair potential configurations,
which greatly simplified the analysis in the MS theory \cite{mak91},\cite
{ste92} , replaces the many electron correlation function with a product of
pair correlation functions, and so greatly overestimates the QP 'scattering'
effect in the asymptotic limit $n_F^{1/2}\gg 1$ . In fact, up to the second
order in $\tilde{\Delta}_0$, our result (Eq.(\ref{p2.16})) is identical to
that obtained by MS (Eq.(\ref{p1.1})). The higher order terms, however,
differ substantially; our quartic term is $1/n_F^{1/2}\ll 1$ smaller than
that obtained by expanding the exponential in Eq.((\ref{p1.1})) up to second
order in $\lambda(\Delta _0,n_F)$.

This result reflects a very interesting phenomenon: In the ground Landau
level approximation for the condensate of Cooper-pairs, the quadratic term
in the free energy expansion is known \cite{man92} to be completely
independent of the vortex lines distribution. Therefore, it has nothing to
do with the broadening of the Landau levels by the inhomogeneous pair
potential in the vortex state. Indeed, in the standard expression \cite
{agd63} for the SC free energy in terms of the 'dressed' electron Green's
function (or the QP Green's function) the entire series of self energy
corrections is multiplied by a second order factor in $\Delta $.
Consequently, the quartic term is the lowest order correction to the free
energy, which contains the 'scattering' effect. It may be, therefore,
concluded that in the asymptotic limit of the 2D model used here, the
'scattering' effect is much weaker than what predicted by any theory
consistent with the random vortex lattice approximation\cite{ste92}\cite
{mak91} \cite{bruun97}.

The structure of our expression for the free energy ,(\ref{fs}),as well as
for the oscillatory magnetization, (\ref{p2.16}), suggests that the small
expansion parameter in the theory is $x\equiv \frac{\widetilde{\Delta }_0^2}{%
n_F}$ rather than $\frac{\widetilde{\Delta }_0^2}{n_F^{1/2}}$, as suggested
by Eq.(\ref{p1.1}). The full expansion should therefore read: 
\begin{equation}
\widetilde{M}_{osc}\approx \widetilde{M}_{n,osc}\left[ 1-\pi ^{3/2}\sqrt{n_F}%
x\Theta (x)\right]  \label{p4.1}
\end{equation}
where at $x\ll 1$ the function $\Theta \left( x\right) $ has an expansion $%
\Theta \left( x\right) \approx 1-\sqrt{2}\beta _Ax$.

Now the expression within the square brackets in Eq.(\ref{p4.1}) vanishes at 
$x\Theta(x)= 1/\pi^{3/2}n_F^{1/2}$. Thus a sign inversion of the magnetic
oscillations amplitude takes place at $x\approx 1/\pi^{3/2}n_F^{1/2}\ll 1$,
where $\Theta(x)\approx 1$, i.e. well within the range of validity of our
expansion \cite{nor95,man96}.

One therefore expects that in a 2D superconductor the dHvA amplitude will
reverse sign due to pairing at a certain field ,$H_{inv}$, below $H_{c2}$,
and remains virtually undamped well below the point of inversion.

This conclusion may be changed if disorder in the vortex lattice, or vortex
lines motion will be taken into account, as indicated by the MS result.
However, application of the MS model to real disordered vortex lattices
should be considered very cautiously since the effect of disorder has not
been introduced self consistently there.

The crossover to the low temperature power law behavior, obtained by Dukan
et al. , is reflected in our theory by the breakdown of perturbation theory
at very low temperature. At such low temperatures, the LK thermal smearing
parameter $X\ll 1$, and our expansion does not exist for all magnetic fields
since the density of states parameter $q_\nu $ diverges like: 
\begin{equation}
q_\nu =\frac{2}{\left\{ 1-\exp \left[ \left( 2\nu +1\right) X\right]
\right\} }\sim \frac{2}{\left( 2\nu +1\right) ^2X^2}  \label{p4.2}
\end{equation}
when a Landau level crosses the Fermi energy with half integer filling
factor $n_F$.

Under this condition, and for sufficiently small $\Delta_{0}$, the SC
pairing is restricted to a single Landau level,and the QP energies are close
to the diagonal elements of the BdG Hamiltonian in the Landau levels
representation, i.e. \cite{duk94},\cite{nor95}: 
\begin{equation}
E_{\overrightarrow{k},n}=\sqrt{\left[ \hbar \omega _c\left( n+1/2-n_F\right)
\right] ^2+\left| \Delta _{n,n}\left( \overrightarrow{k}\right) \right| ^2}
\label{p4.3}
\end{equation}
which is not an analytical function of $\Delta _0^2$ at the Fermi surface.
This also explains the linear dependence of $\lambda(\Delta _0,n_F)$ on $%
\Delta _0$, obtained by Norman et al. \cite{nor95} for small $\tilde{\Delta}%
_{0}$ at low temperatures.

It is interesting to note that in our expansion the quadratic and the
quartic terms for each Matzubara frequency $\nu $ are proportional to $q_\nu 
$ and $q_\nu ^2$ respectively. Thus, the expansion parameter is actually $%
x\sim \frac{\Delta _0^2}{\left( \hbar \omega _c\right) ^2n_F}q_0$

In the high temperature limit $X\geq 1$, where $q_0\approx 2$, it reduces to
the temperature independent value $x\sim \tilde \Delta _0^2/n_F$ used above.
In the very low temperature limit,$X\ll 1$, it diverges with $(1/T)^2$, i.e. 
$x\sim [\Delta _0/\pi k_BT]^2/n_F$

The breakdown of the small $\Delta $ expansion, resulting from this
divergence at sufficiently low temperatures, seems to be related to the
emergence of an opposite, high $\Delta $ expansion in the small parameter $%
\frac{1}{x}\sim (\pi k_{B}T/\Delta_{0})^2$, as obtained by Dukan and
Tesanovic \cite{duk95} (see Eq.(\ref{p1.4})) in the low temperature limit.

Application of the theory developed in the present paper to real
experimental situations is not a straightforward matter; in addition to the
influence of disorder in the vortex lattice and vortex line fluctuations on
the QP 'scattering' , discussed above, the 3D nature of the single electron
band structure could also play an important role . The importance of the
latter effect may be appreciated by noting that in contrast to the 2D model
studied here, in a 3D electron system , e.g. with a spherical Fermi surface
, Cooper-pairs in low Landau levels (i.e. for $n,n^{^{\prime }}\approx 0$ )
and with large longitudinal momenta ( i.e. near $k_z=k_z^{^{\prime }}=k_F$),
have the largest contribution to the SC condensation energy. This region is
far away from the extremal orbit $k_z=0$, $n=n_F$,which dominates the dHvA
oscillations .

As a result, in addition to the QP  near the extremal orbit, their
counterparts with small cyclotron orbits (i.e. for $n\ll n_F$) and large
longitudinal momenta $k_z$ , should also contribute significantly to the SC
free energy in this case. The relatively strong sensitivty of QP with small
cyclotron orbits to 'scattering' by the vortex lattice, as implied by the
large damping parameter $\lambda $ found in Ref..\cite{nor95} , may indicate
that the QP 'scattering' effect in 3D systems is stronger than in the
equivalent 2D systems.

An effective parameter , $n_F$ $^{*}\leq n_F$ , may be therefore introduced
to take into account such an increase in the QP 'scattering'
effect.

Most of the SC materials in which clear dHvA oscillations were observed in
the vortex state, such as $V_3Si$, $Nb_3Sn$, $YNi_2B_2$,and $NbSe_2$, are
essentially 3D systems with complex band structures and nonspherical
Fermi-surfaces. One therefore expects characteristic values of $n_F^{*}$
smaller than $n_F$ in these materials. Furthermore, the nonspherical Fermi
surfaces , combined with some unavoidable deviations from
perfect crystaline order , should lead to some finite distribution around
each dHvA frequency.

A careful examination of the Fourier transformed spectra for these materials 
\cite{cor294,har94,goll96} indeed shows a significantly broad frequency
distribution about each dHvA frequency $F$ (=$Hn_F$ for a spherical Fermi
surface), with line width of the order $\Delta F\sim \left( 100-200\right)
\,T$. This should be compared to the effective range of frequency modulation 
$\left( \Delta F\right) _{inv}=H_{c2}H_{inv}/4\left( H_{c2}-H_{inv}\right) $%
, associated with the expression within the square brackets in Eq.(\ref{p4.1}%
), which does not exceed $15\,T$. Thus it is not surprising that the
measured signal, which is the Fourier transform of this broad spectrum, does
not exhibit a fine structure like the sign inversion  predicted in our ideal
2D electron gas model.

The organic superconductor $\kappa -\left( ET\right) _2Cu\left( NCS\right)
_2 $ seems at first sight a good candidate for testing the predictions of
our theory, due to the quasi 2D nature of its electron band structure .
Unfortunately, the transition from the normal to the SC state observed
experimentally in this material is very broad \cite{wel95}, extending far
below the estimated value of $H_{inv}$,which is found to be very close to
the mean field value of $H_{c2}$ in this material. This is not 
surprising since the low dimensional nature of this compound and the 
low temperatures used in the dHvA experiments can lead to strong quantum 
flucuations in the phase of the order parameter \cite{fisher86},
\cite{spivak95}, and so to the breakdown of the mean field approximation 
used in our theory.

The relatively weak QP 'scattering' , predicted in the present paper, 
seems to be confirmed, however, by the majority of the experiments
performed so far:  According to our theory it should lead to a significant 
deviation of the experimentally measured amplitude from
the Maki-Stephen-Wasserman fitting formula (see Eq.(\ref{p1.1},\ref{p1.2}))
in the region where the leading SC effect exceeds the zeroth order (i.e.
normal electron) term, i.e for $H\geq H_{inv}$. In this region the above
qualitative analysis indicates that the damping of the dHvA oscillations may
be described by a parameter , $\widetilde{\Delta }_0^2/n_F$ $^{*}$ , smaller
than the characteristic MS parameter $\lambda \sim \widetilde{\Delta }_0^2/%
\sqrt{n_F}$ . Such a crossover from a relatively strong damping just below $%
H_{c2}$, described well by the MS fitting formula, to a weaker damping at
lower fields, was indeed observed in almost all experiments carried 
out so far\cite{cor294,har94,goll96}.

Furthermore, from the available experimental data two different characteristic
slopes of the corresponding Dingle plot can be clearly distinguished . Our
estimations show that the experimental crossover field $H_{cross}$ from one
slope to another is in a good agreement with the calculated inversion field $%
H_{inv}$. In particular, we have obtained for $V_3Si(F=1570T)$: $%
H_{cross}\sim 12.5T$, $H_{inv}\sim 13.8T$, for $YNi_2B_2(F=511T)$: $%
H_{cross}\sim 4.5T$, $H_{inv}\sim 4T$, for $NbSe_2(F=152T)$: $H_{cross}\sim
5.6T$, $H_{inv}\sim 4T$, and for $Nb_3Sn(F=581T)$: $H_{cross}\sim 11.4T$, $%
H_{inv}\sim 13.7T$. 

{\bf ACKNOWLEDGMENT}: We acknowledge valuable discussions with G.M. Bruun,
S.Hayden, W. Joss, A. MacDonald, M.R. Norman , E. Steep , B. Spivak and Z.
Tesanovic. This research was supported by a grant from the US-Israeli
Binational Science Foundation grant no. 94-00243, by the fund for the
promotion of research at the Technion, and by the center for Absorption in
Science, Ministry of Immigrant Absorption State of Israel.

\end{document}